# Crystal rotation during alloy heating

Zhida Liang[1,2,*], Emad Maawad[1], Florian Pyczak[1]

1. Institute of Materials Physics, Helmholtz-Zentrum Hereon, Max-Planck-Strasse 1, Geesthacht 21502, Germany
2. Department of Materials Science and Metallurgy, Cambridge University, 27 Charles Babbage Rd, Cambridge CB3 0FS, UK

∗ Corresponding author: Zhida Liang, zhida.liang@outlook.com

**Abstract**

**Over past decades, high energy X-ray diffraction techniques at synchrotron beamline offer us the unique possibility to follow concomitantly some mechanisms on the grain level in metallic materials. Crystal rotation, as one important factor for the crystal orientation distribution, usually influences the anisotropic properties of materials[1-3]. Crystal rotation has been visualized in deformation[4], recrystallization[5], and chemical reaction processes[6]. In this study, we firstly visualized an unexpected crystal rotation during heating up of superalloys as well by in situ 3DXRD scanning microscopy which has not been previously reported yet. The crystal rotation was found to be related to the locked-in first order residual strain relief during the annealing treatment. In addition, intermittent dynamic, *i.e.*, some individual diffraction spots from Debye-Scherrer rings appearing, disappearing, and reappearing, occurred in this study which has ever been thought to be caused by subgrain formation in deformation experiments[4]. *However*, by residual strain analysis coupled with finite element simulation, we found that the intermittent dynamic is married with thermal expansion, crystal distortion, and crystal rotation as well. This finding opens a new world for deeply understanding the residual strain relief during materials, particularly metal, heating.**

**Introduction**

Crystal rotation, a common occurrence in materials science during deformation[4], recrystallization[5], and chemical reaction processes[6], plays a pivotal role in determining the crystal orientation distribution, *i.e.*, texture, influencing the anisotropy of the materials' mechanical properties[1, 2]. Understanding crystal rotation contributes significantly to comprehending dislocation movement and reorientation of metallic crystal. Therefore, exploring the interplay between lattice orientation, deformation, and thermal effects provides us a valuable insight into the intricate behavior of engineering materials.

Intermittent dynamic phenomena, *i.e.*, some individual diffraction spots from Debye-Scherrer rings appearing, disappearing and reappearing, was proposed to occur during proceeding deformation due to subgrain formation which has been demonstrated by three-dimensional x-ray diffraction (3DXRD) scanning microscopy[4]. In the recrystallization process, usually, new individual diffraction spots appear from Debye-Scherrer rings during annealing treatment assuming that it corresponds to a newly recrystallized grain[7]. Nevertheless, the metal deformation and recrystallization are usually accompanied with crystal rotation[3] not only subgrain formation.

In this study, we found the intermittent dynamic of diffraction spots during the annealing treatment of a polycrystalline superalloy with cast condition, as is shown in **Fig. 1**. By employing a single crystal with the same composition, we investigated the interplay between intermittent dynamic, crystal rotation and residual elastic deformation. To enable such measurements, we established a dedicated in situ heating 3DXRD setup based on a hexapod, illustrated in **Fig. 1(A)**, to capture the dynamic crystal rotation behavior beamline P07, operated by Hereon at the storage ring PETRA III at Deutsches Elektronen-Synchrotron (DESY). The experimental setup consists of a cylinder sample, a high temperature vacuum system, a cooling system and an in situ high energy diffraction system (a highly penetrating 100keV X-ray beam which cross-section is 1×1 mm$^2$), as illustrated in **Fig. S3**. We revealed that the residual stress relief caused by annealing treatment is the culprit to govern crystal rotation which was not reported before.

**Results and discussion**

**Intermittent dynamic during heating up revealed by in situ 3DXRD**

An in situ annealing treatment was applied to a polycrystalline CoNiCr-3W superalloy sample (5 mm in diameter and 7.5 mm in length) with binary γ/γ′ phases (refer to **Fig. S1(A)**). To maximize the effective volume irradiated by X-rays, the sample was oscillated angularly, allowing X-rays scanning with ± 40° (φ) around the y-axis in laboratory coordinates. The Debye-Scherrer rings, as shown in **Fig. 1(B)**, were recorded after the temperature stabilized at the pre-set value for approximately 10 min. A section of Debye-Scherrer ring (002) was unfolded for better clarity, and eight diffraction spots were tracked to investigate the intermittent dynamic evolution in **Fig. 1(C)**. The results show the intermittent dynamic phenomena that occurred throughout the annealing at different temperatures. For example, spots 1 and 4 disappear at 600 °C and reappear at 900 °C. Spot 7 nearly disappears at 300 °C (its intensity is very low according to its line profile, marked by a black circle in **Fig. 1(D)**), reappears at 600 °C, and disappears again at 900 °C. In addition, 2 new diffraction spots appear at 600 °C but disappear again at 900 °C, marked by red circle in **Fig. 1(C)** and **(D)**. According to the textbooks, the intersection of reciprocal lattice points and the Ewald sphere decreases due to thermal expansion as the temperature rises which can lead to the disappearance of diffraction spots. This is a plausible explanation. In addition, the intensity of spots is influenced by atomic vibrations at elevated temperatures, but such vibrations alone cannot account for spots disappearing, reappearing, and then disappearing again. This kind of intermittent dynamic was previously thought to be linked with subgrain formation[4].

However, our investigation challenges this notion. Since our sample is in a cast condition without undergoing high plastic deformation and therefore there is not enough driving force to drive subgrain formation through recrystallization. We propose that elastic strain and crystal rotation occurring during in situ heating up have a high possibility of resulting in the observed intermittent dynamic.

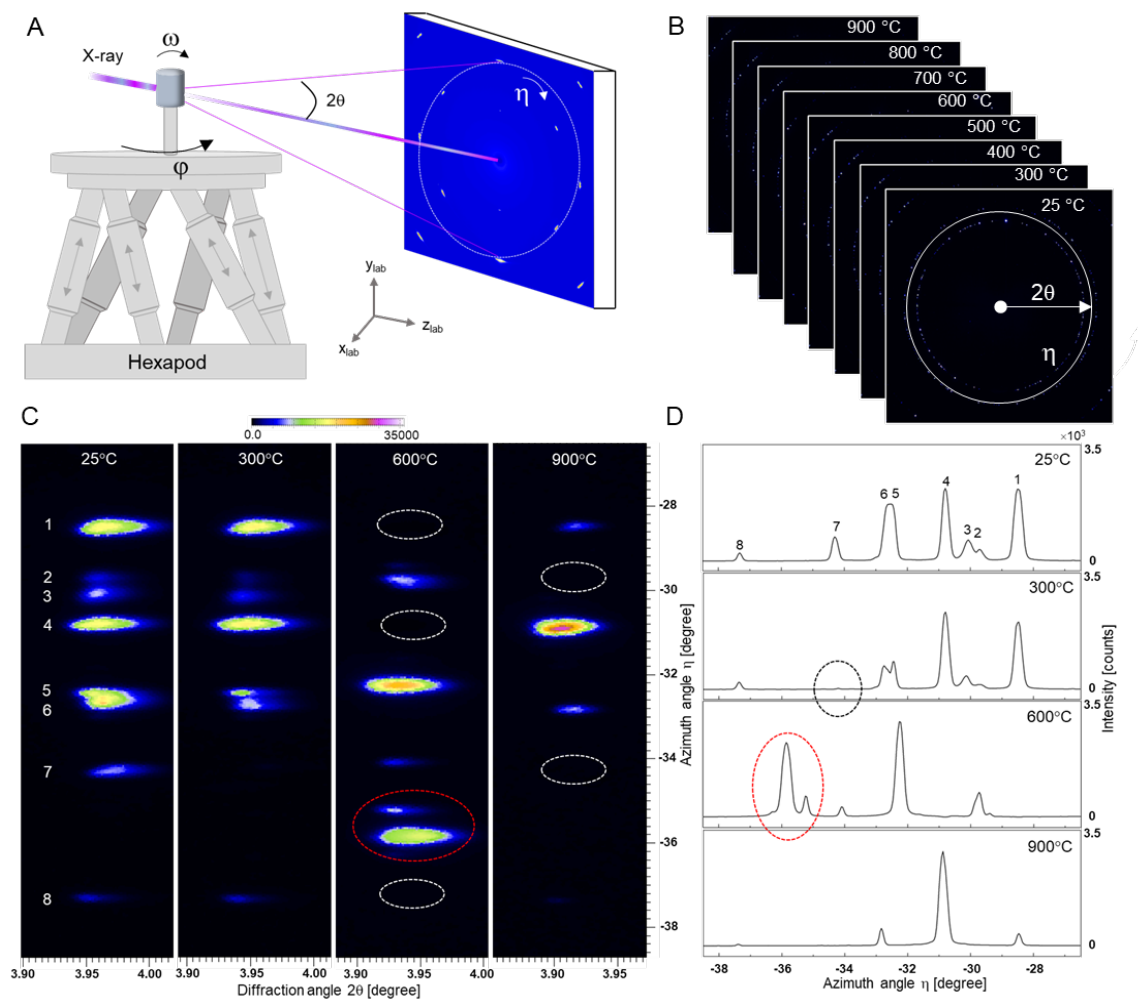

**Fig. 1 In situ 3DXRD setup based on Hexapod.** (A) Sketch of the 3DXRD scanning experimental setup based on hexapod, allowing the sample to be tilted and rotated. The coordinate system ($x_{ab}$, $y_{lab}$, $z_{lab}$) and angles ($\omega$, $\eta$, $\theta$, $\varphi$) are defined. (B) The Debye-Scherrer rings during in situ heating up of the CoNiCr-based polycrystalline superalloy (CoNiCr-3W) with ± 40° 3DXRD scanning ($\varphi$). (C) The lay-out view of Debye-

Scherrer rings and (D) the resultant integrated line profiles of diffraction spots (3.7° ≤ 2θ ≤ 4.1°) both of which show direct evidence of intermittent dynamic that some individual diffraction spots from Debye-Scherrer rings disappear, reappear, and disappear again during annealing treatment (the white circles represent disappearing of spots; the red circles represent appearing of new spots).

**Direct observation of crystal rotation in dependence of temperature**

To observe crystal rotation, easily track the variation of single diffraction spot and exclude the effect of recrystallization during annealing treatment, a self-developed high quality single crystal CoNiCr-3W superalloy with the same composition Co-35Ni-15Cr-5Al-5Ti-3W (at.%) fabricated by Bridgman casting system (**Fig. S1(B)**) was employed. In **Fig. 2(A)**, the typical high-quality single crystal diffraction spots were obtained through 3DXRD oscillation scanning with ±180° (φ). A single spot at an azimuth angle of 90° was fixed to track crystal rotation as the temperature varied. For mosaic-structured materials, the rocking angle (ω) influences the intensity distribution along the η angles on the two-dimensional detector images. At each temperature, 301 images were recorded while stepping from –3° to +3° in Δω, around the position Δω = 0°, where the reflection spot (002) is the brightest at the room temperature. To eliminate the effect of mosaicity and ensure a fair comparison between reflection spots at different temperatures, the summed reflections from all rocking angles and azimuth angles were integrated, as shown in **Fig. 2(B)** and **(C)**.

In-situ measurements were conducted at various temperatures ranging from 25 °C to 1050 °C. The temperature-dependent evolution of diffraction spots along rocking angles and azimuth angles is given in **Fig. 2(D)** and **(F)**. As the sample temperature changes, the spots exhibit a shift towards a smaller diffraction vector modulus, denoted as $|\vec{g}|$. Besides, the spots undergo movement along different ω and η angles. The spatial positions of features, characterized by the (θ, ω, η) distributions from 3DXRD rocking scanning, were determined using the center-of-mass (COM) algorithm, similar to methodologies presented in literatures[8,9].

With the temperature increasing, there is one 'elbow' appearing in **Fig. 2(D)** and **(F)**, respectively. Based on line profiles in **Fig. 2(E)** and **(G)**, it is evident that this 'elbow' occurred at 900 °C, corresponding to sample aging heat treatment temperature. The shift of observed patterns along the rocking angle ω and azimuth angle η (forming oblique lines) indicates that an interesting "crystal rotation" behavior occurs during the annealing treatment of the sample.

The pole figure provides a direct visualization of crystallographic rotations (for transformation formulas from the spatial positions of diffraction spots to the pole figure, refer to Materials and Methods in the supplementary materials). In the pole figure (**Fig. 2(H)**), the lattice plane (001) undergoes a rotation from (001) towards (011) between 25 °C and 900 °C, followed by a slight rotation back from 900 °C to 1050 °C. To quantify the magnitude of the lattice rotation angle, the misorientation between different temperatures was calculated based on Euler angles ($\phi$, $\varphi_1$, $\varphi_2$) as outlined in **Table S1,** with temperature 900 °C as a reference case. As the temperature increases, the misorientation deceases from room temperature until it reaches zero at 900 °C.

Beyond this temperature, the misorientation difference among crystals at various temperatures expands. This indicates a kind of manufacturing operation taking place at 900 °C, specifically fast cooling after the aging heat treatment (annealing treatment at 900 °C for 200 h), which introduces residual locked-in strain into the sample. Upon reheating, the release of residual elastic strain energy coincides with crystal rotation and the subsequent wrapping of the sample. To confirm relationship of aging heat treatment temperature and crystal rotation 'elbow' repeatably, another single crystal CoNiCr-baseline, which experienced annealing treatment at 800 °C for 200 h followed by air quenching, was used to be tested by in situ heating as well. Confirmatively, the 'elbow' occurred at 800 °C again, corresponding to sample aging heat treatment temperature, which was shown in **Fig. S4**.

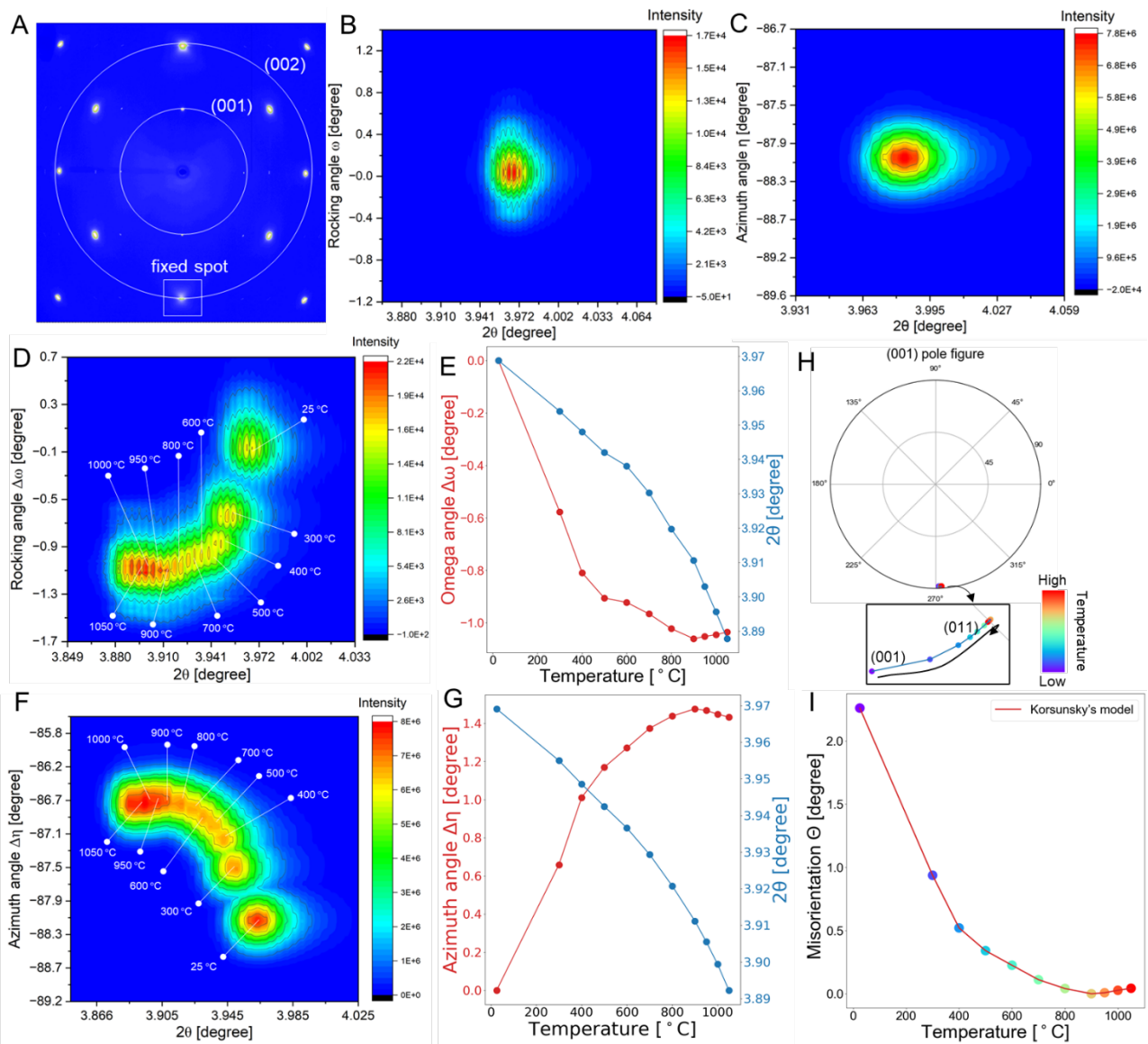

**Fig. 2 Direct observation of crystal rotation.** (A) The diffraction spots obtained through 3DXRD with a ±180° scanning (φ), representing a high-quality single crystal (CoNiCr-3W). (B) The contour map of reflection spots (002) with varying rocking angles (ω). (C) The contour map of reflection spots (002) with different

azimuth angles (η). (D) The rocking angle (Δω) and diffraction angle (2θ) diffraction spots (002) at different temperatures. (E) Spatial positions for rocking angle (Δω) and diffraction angle (2θ) of the spots based on the center-of-mass (COM) method, singled out in **Fig. 2(D)**. (F) The summation of azimuth angle (Δη) and diffraction angle (2θ) reflection spots at different temperatures. (G) Spatial positions for azimuth angle (Δη) and diffraction angle (2θ) of the reflection based on the COM method, singled out in **Fig. 2(F)**. (H) The stereographic projection pole figure of crystal at different temperatures, based on Korsunsky's model[10]. (I) The misorientation of the single crystal at different temperatures (the temperature 900 °C as reference case).

## The origin of lattice rotation from the first order residual strain

To identify the existence of the first-order residual elastic strain in the sample, the Debye-Scherrer rings obtained from **Fig. 1(B)** were applied for residual strain analysis roughly. The analysis of full diffraction rings enables the simultaneous determination of all strain components in the plane. If there is residual strain existing in samples, the Debye-Scherrer rings change from a circle into an ellipse, illustrated in **Fig. 3(A)**. **Fig. 3(B)** presents one diffraction pattern of the polycrystalline CoNiCr-3W alloy at room temperature.

The (002) Debye-Scherrer rings collected at azimuthal angles η of 0° and 90° were used to evaluate lattice parameters $d_a$ and $d_t$ along the cylinder's axial direction and tangential direction along the beam path, respectively.

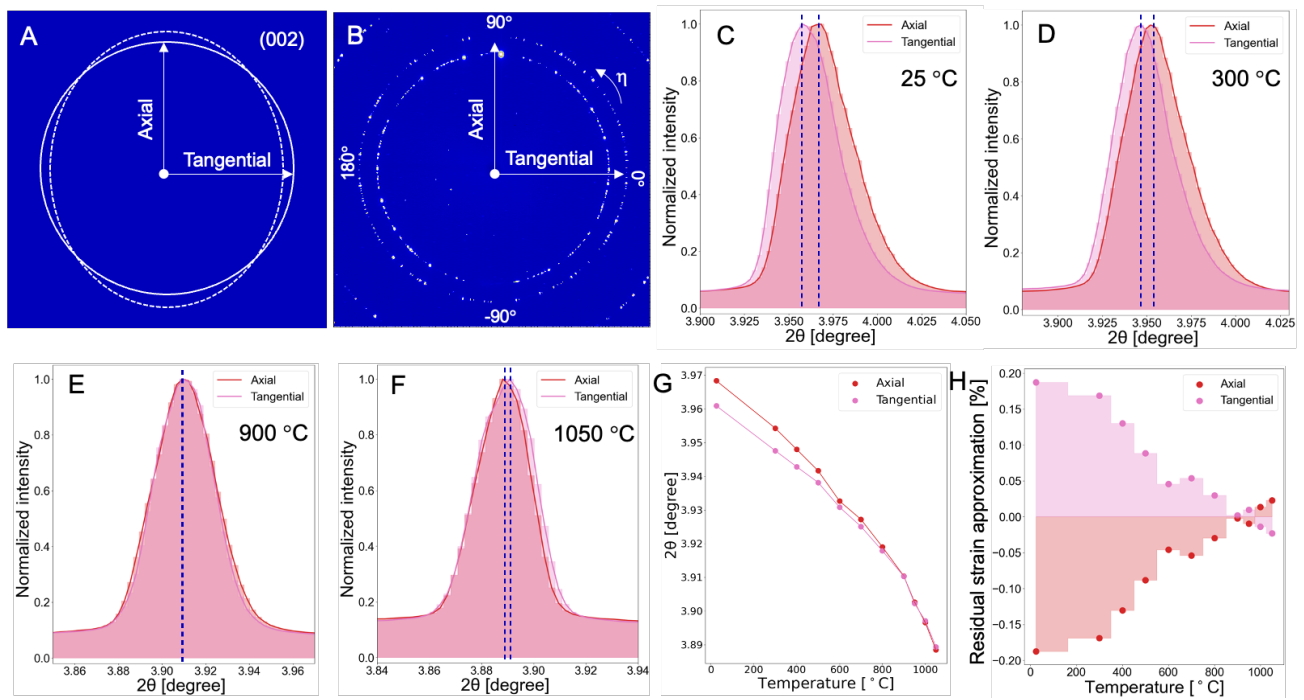

**Fig. 3 The first-order residual strain analysis**. (A) Schematic diagram of Debye-Scherrer rings with and without residual strain (with residual elastic strain, the Debye-Scherrer ring becomes an ellipse.). (B) The Debye-Scherrer rings of the CoNiCr-based polycrystalline superalloy (CoNiCr-3W) at room temperature from **Fig. 1(B)**. (C)-(F) The line profiles of the intensity as function of the diffraction angle (2θ) at different

temperatures in axial and tangential directions, respectively. (G) The diffraction angle (2θ) at different temperatures in axial and tangential directions and (H) the residual strain approximation in the axial and tangential directions.

The selected spots at azimuthal angles η of 0° and 90° were integrated into line profiles for comparison, as depicted in **Fig. 3(C)-(F)** for selected temperatures and **Fig. S5**. As temperature rise, the diffraction angle (2θ) difference become smaller and it is almost zero at 900 °C, see **Fig. 3(G) and Fig. S5(H)**. Hence, the two-dimensional distributions of axial and tangential residual elastic strains were approximately calculated by $\varepsilon_a = 2(d_a - d_t)/(d_a + d_t)$ and $\varepsilon_t = 2(d_t - d_a)/(d_a + d_t)$. Actually, the residual strain in a real object is often asymmetric due to geometry shape[11]. However, the results in **Fig. 3(H)** can reflect the approximation of crystal extension or contraction in an object to some extent. Based on the calculated residual strain approximation results, the axial and tangential strain conditions of the sample can reflect tensile and compression strain, respectively. With temperature rising, the resultant strain decreases gradually until it almost reaches zero at 900 °C. Above 900 °C, the strain changes slightly. The turning point in strain variation also occurs at 900 °C which situation is consistent with crystal rotation in **Fig. 2(I)**. Therefore, we postulate that the crystal rotation is closely associated with residual strain relief induced by the annealing treatment.

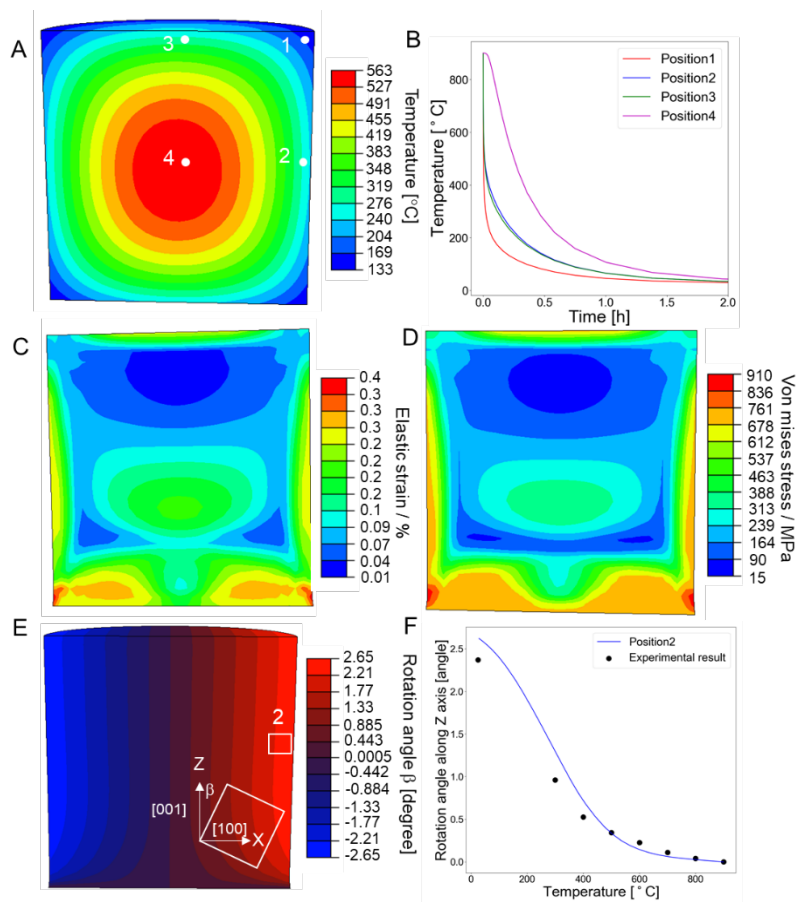

**Fig. 4 Thermomechanical finite element simulation results.** (A) The temperature distribution of a cylinder sample during air quenching from 900 °C (the preset temperature of cylinder is 900 °C). (B) The temperature

evolution of 4 positions from **Fig. 4(A)** during air quenching. (C) Elastic strain and (D) Von Mises stress distribution at room temperature after air quenching. (E) Thermomechanical simulated rotation angle long Z-X plane at room temperature. (F) Comparison of the evolution of simulated rotation angles long Z-X plane and the experimentally observed rotation angles from **Fig. 2 (I)**.

The thermomechanical finite element method (FEM) was employed to simulate the temperature field and the first-order residual locked-in strain and stress (for details see supplementary materials). **Fig.4 (A)** and **(B)** presents the temperature distribution and different cooling speed at different positions of cylinder throughout air quenching. In the simulation results, we were inspired that the temperature gradients in the cylinder during fast cool-down are the main reason for the first-order residual strain and stress formation, as illustrated in **Fig.4 (C) and (D)**. From FEM simulation in **Fig. 4(D)**, we can see that the Von mises stress in some regions exceed the yield stress known from experimental results as listed in **Table S2**. Accordingly, the FEM simulation predict plastic deformation taking place in a small volume at the wall of the cylinder during air quenching. When we performed reheating experiments, the residual strain was released gradually verified by XRD analysis results in **Fig.3 (H)**. The residual stress relief is commonly accompanied with elastic strain recovery[12,13]. Based on the theory of solid mechanics, the deformation gradient **F** can be expressed by **F=RU,** where **R** is the rotation tensor for describing the rigid-body rotation and **U** is a symmetric tensor[14]. As a result, crystal rotation generally occurs in metallic materials when there is deformation. The sample rotation data from position 2, shown in **Fig. 4(E)**, was extracted as a function of temperature from the FEM simulation for comparison with the experimental results, see **Fig. 4(F)**. With the temperature decreasing, the rotation angle from the simulation decreases which is compatible with experimental results from **Fig. 2(I)**.

**Discussion and summary**

In our investigation, we found crystal rotation to be related to residual strain relief during annealing treatments. The residual strain resulted from the temperature inhomogeneity during the sample air quenching process. *However, is crystal rotation the primary origin for the intermittent dynamic of X-ray diffraction spots?* Indeed, crystal rotation is not the only symptom observed during the annealing treatment, there are also two other notable events, *i.e.*, crystal distortion recovery and thermal expansion. To elucidate these relationships, we illustrated the effects of these three different mechanisms in reciprocal space, as illustrated in **Fig. 5(A)**. **i**: With the temperature increasing, the crystal expands, and thus the reciprocal point moves along the direction of the diffraction vector $\vec{g}$ so that the intersection of the reciprocal lattice point and Ewald sphere becomes smaller, bigger, or disappears at high temperatures. However, this mechanism has a very small influence on the intermittent dynamic for a small-wavelength X-ray beam ($\lambda$ = 0.124 Å), especially at the high-energy X-ray beamline based on synchrotron radiation, as the arc of the Ewald sphere is almost flat. **ii**: Based on rocking curves and their full width at half maximum (FWHM) statistics depicted in **Fig. 5(B)** and **(C)**, the mosaicity decline becomes evident as the temperature rises to 900 °C, indicating a reduction in crystal distortion magnitude. Moreover, the findings in **Fig. 3** demonstrate that temperature variations can induce either crystal extension or contraction through residual stress relief, significantly impacting the intersection of reciprocal lattice points and the Ewald sphere. According to **Fig. 3(C)-(F)** and **Fig. S5(H)**, the maximum angle (2θ) difference in diffraction vector $\vec{g}$ direction was observed at room temperature, which is 0.007°. **iii**: **Fig. 2(H)** shows that the crystal rotates from the (001) plane towards the (011) plane, and in **Fig. 2(I)**, the maximum rotation magnitude is approximately 2.5°. Consequently, the crystal rotation presents a higher likelihood of causing reciprocal lattice points to move out from the Ewald sphere. However, the sample in this study encountered not just one of the aforementioned possibilities, but rather a combination of all of them. Because the first order residual stress relief, occurring during the annealing treatment, leads to simultaneous processes of crystal rotation, crystal distortion recovery, and thermal expansion.

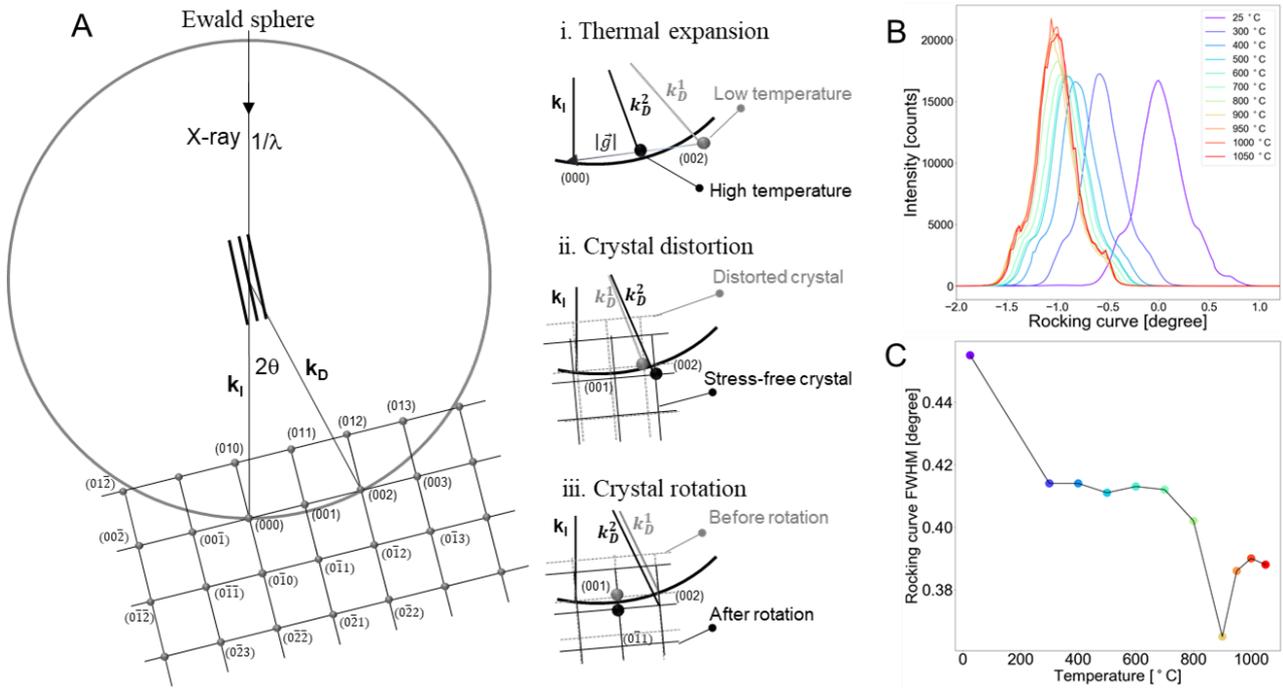

**Fig. 5 Interpretation of intermittent diffraction spot dynamic.** (A) Schematic illustration of the Ewald-sphere construction and three possibilities for the intermittent dynamic - (i) Thermal expansion, (ii) Crystal distortion, and (iii) Crystal rotation. (B) Rocking curves of single crystal CoNiCr-3W at different temperatures. (C) The full width at half maximum (FWHM) of rocking curves from **Fig. 5 (B)**.

Based on the analysis above, intermittent dynamics can take place in various manufacturing processes, including cold rolling, hot forging, heating, and quenching, *etc*. For example, in the cold rolling process[15, 16], crystal distortion, rotation, and subgrain formation are involved. In the hot forging process[17, 18], particularly at 1100 °C in superalloys[19], the intermittent dynamic becomes more complex, encompassing crystal distortion, crystal distortion recovery, crystal rotation, new subgrain formation and evolution of low-angle grains into high-angle grains, *etc*. This study reminds us the intermittent dynamic, as a common phenomenon in X-ray diffraction experiments, can not be triggered solely by subgrain formation. Therefore, the intermittent dynamic should not be simplistically regarded as a signal of recrystallization which provides a new perspective towards X-ray diffraction experiments.

**Acknowledgements**

We thank Prof. Henning Friis Poulsen at Technical University of Denmark, Denmark, on 3DXRD method discussion, Prof. Dagmar Gerthsen at Karlsruhe Institute of Technology, Germany, to review this article and provide productive suggestions, Dr. Peter Staron at Helmholtz-Zentrum Hereon, Germany for residual stress and strain analysis, Dr. Qiang Wang at Central South University, China, on finite element method (FEM) simulation discussion, Mr. Dirk Matthiessen and Mrs Ursula Tietze at Hereon beamline P07 at PETRA III of Deutsches Elektronen-Synchrotron (DESY), Germany, on building the hexapod-based 3DXRD experimental platform.

# Supplementary materials

## Crystal rotation during alloy heating

Zhida Liang[1,2,*], Emad Maawad[1], Florian Pyczak[1]

3. Institute of Materials Physics, Helmholtz-Zentrum Hereon, Max-Planck-Strasse 1, Geesthacht 21502, Germany
4. Department of Materials Science and Metallurgy, Cambridge University, 27 Charles Babbage Rd, Cambridge CB3 0FS, UK

∗ Corresponding author: Zhida Liang, zhida.liang@outlook.com

**Materials and Methods**

**1. Materials and sample preparation**

The self-developed CoNiCr-based polycrystalline[20] (CoNiCr-3W) and single crystal superalloy with binary γ/γ′ phases were used to investigate lattice rotation during heating up in this work. Its chemical composition in atomic percentage is 47Co-35Ni-15Cr-5Al-5Ti-3W. The raw materials were firstly melted to produce buttons by arc melting and then the arc-melted product buttons were used to fabricate single crystal in Bridgman casting system. The experimental sample, CoNiCr-3W, having a gauge length of 7.5 mm and a diameter of 5 mm, experienced a 200 h aging heat treatment at 900 °C and then air quenching, followed by 24 h homogenized heat treatment at 1350 °C and cooling in the furnace. The sample CoNiCr-baseline experienced a 200 h aging heat treatment at 800 °C and then air quenching which was considered as reference sample to investigate annealing temperature effect on lattice rotation. The backscattered-electron scanning electron microscopy image in **Fig. S1(A)** shows the microstructure of the polycrystalline sample (the grain size is about 93 um). The single crystal bar and its microstructure with binary γ/γ′ phases (the particle size is about 120 nm) were shown in **Fig. S1(B)** and **(C)**.

The tested sample (CoNiCr-3W and CoNiCr-baseline) underwent a 360° 3DXRD scanning to assess the quality of single crystal structure through the observation of X-ray diffraction spots. By comparison with another sample containing subgrains (CoNiCr-baseline with a composition of 50Co-35Ni-15Cr-5Al-5Ti, in atomic percent), the sample CoNiCr-3W investigated in this study was demonstrated as a high-quality single crystal structure, depicted in **Fig. S2**.

**2. 3D XRD experiments**

The sample was investigated by high energy X-ray diffraction (HEXRD, $\lambda$ = 0.124 Å) at the synchrotron beamline HEMS run by Helmholtz Zentrum Hereon at the PETRA III storage ring of the Deutsches Elektronen-Synchrotron (DESY, Hamburg, Germany). The beamline operates in the 100 keV with a double crystal monochromator which allows for measuring with complex sample environment while maintaining the possibility to access a wide **q**-range. The sample to detector distance was fixed at 2500 mm to obtain a higher angular resolution at small diffraction angle, ideal for coherent structure characterization and crystallography. The area detector **Perkin-Elmer XRD1621** was employed to record diffraction spots which is an amorphous silicon flat panel detector with CsI scintillator. The dedicated three-dimensional XRD (3DXRD) microscopy station should have high precision x, y and z translations as well as additional rotational degrees of freedom.

In addition, to avoid interference with oxide diffraction information, it is necessary to employ a high vacuum furnace at high temperatures. Therefore, a hexapod was employed into carry high temperature vacuum furnace instead of Eulerian cradle (**Fig. S3**). In high temperature vacuum furnace, there is one additional rotation system that allows sample to be rotated up to 360°. The sample temperature was controlled by a proportional-integral-derivative (PID) controller. The sample heating speed is 20 °C/mins. The in situ heating up experiments were performed at 25 °C, 300 °C, 400 °C, 500 °C, 600 °C, 700 °C, 800 °C, 900 °C, 950 °C, 1000 °C, and 1050 °C, respectively. The thermocouple is S-Type (Platinum for cathode and Platinum 10% Rhodium for anode).

Before initiating in situ experiments, the sample was initially adjusted to its eucentric height to ensure that hexapod tilting or rotation would not impact the sample's position relative to the X-ray beam direction. The sample is affixed to an ω tilting and ϕ rotation stage, where ω represents tilting around an axis perpendicular to the incoming beam, as illustrated in **Fig. 1(A)**.

To fulfil the Bragg condition at room temperature, the single crystal sample is tilted along the angle ω and rotated along the angle φ continuously to find the brightest spot (002) near the azimuth angle η = - 90°. Then, the rocking scanning along angle ω (-3° ~ +3°) was executed with a step of Δω = 0.02°.

### 3. Crystal misorientation and stereographic projection pole figure calculation method

The angles (2θ, η, ω) were extracted from reflection spots by the centre of mass (COM) algorithm which was realized by self-written Python codes.

In the pole figure calculation, we also adopted another transformation procedure which can be expressed by [10]

$$I(2\theta, \eta, \omega) \xrightarrow{transformation} P(a, b) \xrightarrow{projection} P(\alpha_E, \beta_E).$$

when there is no rotation or tilting, the Cartesian coordinates of the an *hkl* pole on the unit sphere are given by

$$r = \begin{bmatrix} \cos(\theta)\cos(\eta) \\ \cos(\theta)\sin(\eta) \\ \sin(\theta) \end{bmatrix}.$$

After rotation or tilting by an angle of ω, the vector is transformed into

$$r' = R \cdot r = \begin{bmatrix} \cos(\omega) & 0 & \sin(\omega) \\ 0 & 1 & 0 \\ -\sin(\omega) & 0 & \cos(\omega) \end{bmatrix} \cdot \begin{bmatrix} \cos(\theta)\cos(\eta) \\ \cos(\theta)\sin(\eta) \\ \sin(\theta) \end{bmatrix}$$

The longitude *a* and latitude *b* on the spherical surface have the following relationship with matrix *r'*

$$\tan(a) = r'(2)/r'(1),$$

$$\tan(b) = r'(3)/\sqrt{r'(1)^2 + r'(2)^2}.$$

This new intensity function $P(a, b)$ is projected onto the sphere's equatorial plane using equal area projection. The relationship between the spherical coordinates $P(a, b)$ and the pole figure coordinates $P(\alpha_E, \beta_E)$ is expressed by

$$\alpha_E = a,$$

$$\beta_E = \sqrt{2}\sin(\pi/2 - b)/2 \times 90°,$$

where $\beta_E$ is the pole distance angle [0, 90°] and $\alpha_E$ is the azimuthal angle [-180°, 180°]. The pole figure was visualized by self-written Python codes.

$$U = \begin{pmatrix} U_{11} & U_{12} & U_{13} \\ U_{21} & U_{22} & U_{23} \\ U_{31} & U_{32} & U_{33} \end{pmatrix} =$$

$$\begin{bmatrix} \cos(\varphi_1)\cos(\varphi_2) & -\cos(\varphi_1)\sin(\varphi_2) & \\ -\sin(\varphi_1)\sin(\varphi_2)\cos(\phi) & -\sin(\varphi_1)\cos(\varphi_2)\cos(\phi) & \sin(\varphi_1)\sin(\phi) \\ \sin(\varphi_1)\cos(\varphi_2) & -\sin(\varphi_1)\sin(\varphi_2) & \\ +\cos(\varphi_1)\sin(\varphi_2)\cos(\phi) & +\cos(\varphi_1)\cos(\varphi_2)\cos(\phi) & -\cos(\varphi_1)\sin(\phi) \\ \sin(\varphi_2)\sin(\phi) & \cos(\varphi_2)\sin(\phi) & \cos(\phi) \end{bmatrix}.$$

The Bunge orientation matrix *g* is the orientation matrix *U* transpose[21,22], i.e.,

$$g = U^T.$$

The orientation relationship between the two crystals, *e.g.*, crystals 1 and 2, can be described by the equation

$$\Delta g_{12} = g_1 \cdot g_2^{-1} = \begin{pmatrix} g_{11} & g_{12} & g_{13} \\ g_{21} & g_{22} & g_{23} \\ g_{31} & g_{32} & g_{33} \end{pmatrix},$$

where $g_2^{-1}$ is the inverse matrix of $g_2$.

The misorientation angle Θ between two crystals can be defined by

$$\Theta = \arccos\left(\frac{g_{11}+g_{11}+g_{11}-1}{2}\right).$$

The stereographic projection figures, *i.e.*, pole figure, were visualised by self-written python codes.

## 4. Structure characterization

The microstructures were characterized in the backscattered electron (BSE) mode in a scanning electron microscope (FE-SEM Gemini, Zeiss, Germany) at 15 keV electron energy.

## 5. Finite element method simulation

The air cooling of the cylinder (5 mm diameter and 7.5 mm length, referring to **Fig. S6(A)**) from 900 °C to 25 °C was simulated by the finite element method (FEM) executed in Abaqus/CAE 2023[24]. The alloy CoNiCr-3W used in this study was modelled as a homogeneous isotropic elasto-plastic material. The simulation parameters required, such as Young's modulus[25], Poisson ratio[26], heat conductivity[27], density[20] (8570 kg/m³), thermal expansion coefficient[27] and yield stress[20] were listed in **Table S2**. Young's modulus as a function of temperature (**T**) was calculated based on expression[28, 29]: **E** = 2(1+*v*)×(87.32-0.0009**T**-0.000019**T**²). The bottom plane of the cylinder was constrained as mechanical boundary condition for stress and strain model simulation, referring to **Fig. S6(B)**. The cooling interaction was applied to all surface of cylinder, as illustrated in **Fig. S6(C)**. The radiative heat loss is represented by Newton's cooling law with a surface emissivity of 0.7.

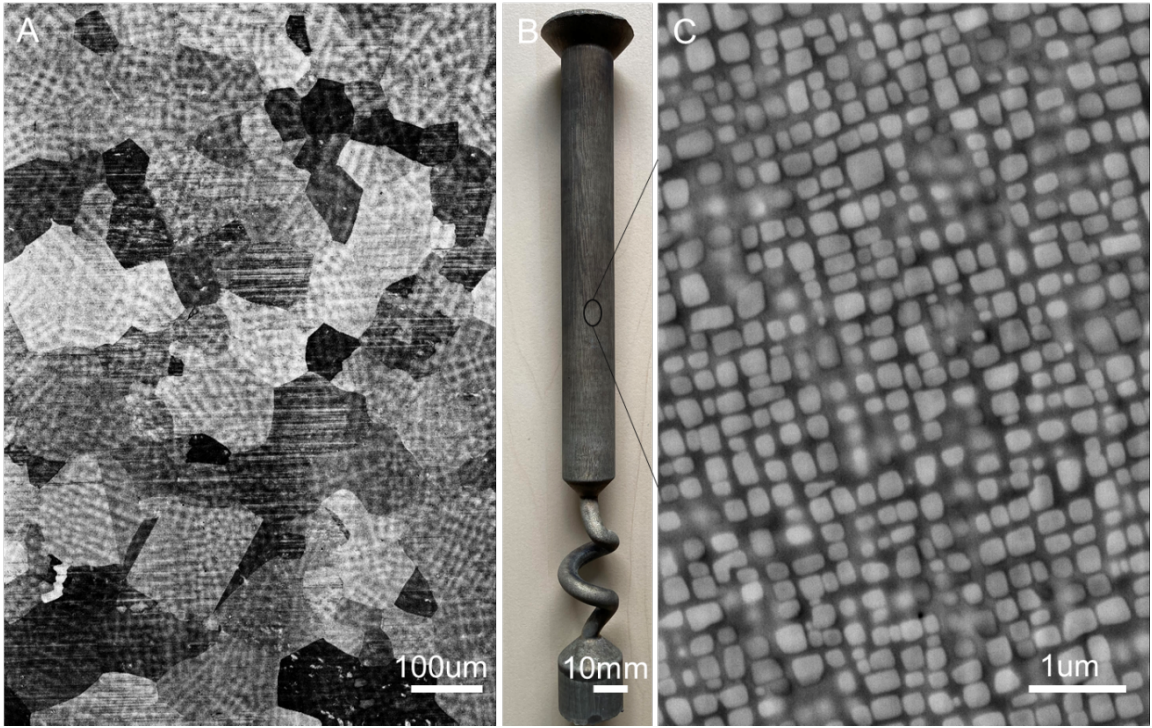

**Fig. S1** (A) The microstructure of CoNiCr-based polycrystalline superalloy (CoNiCr-3W) with composition Co-35Ni-15Cr-5Al-5Ti-3W (at.%). (B) The in situ heating CoNiCr-based single crystal superalloy fabricated by Bridgman casting system which composition is same to polycrystalline CoNiCr-3W. (B) The microstructure with binary γ′/γ phases of the CoNiCr-based single crystal superalloy.

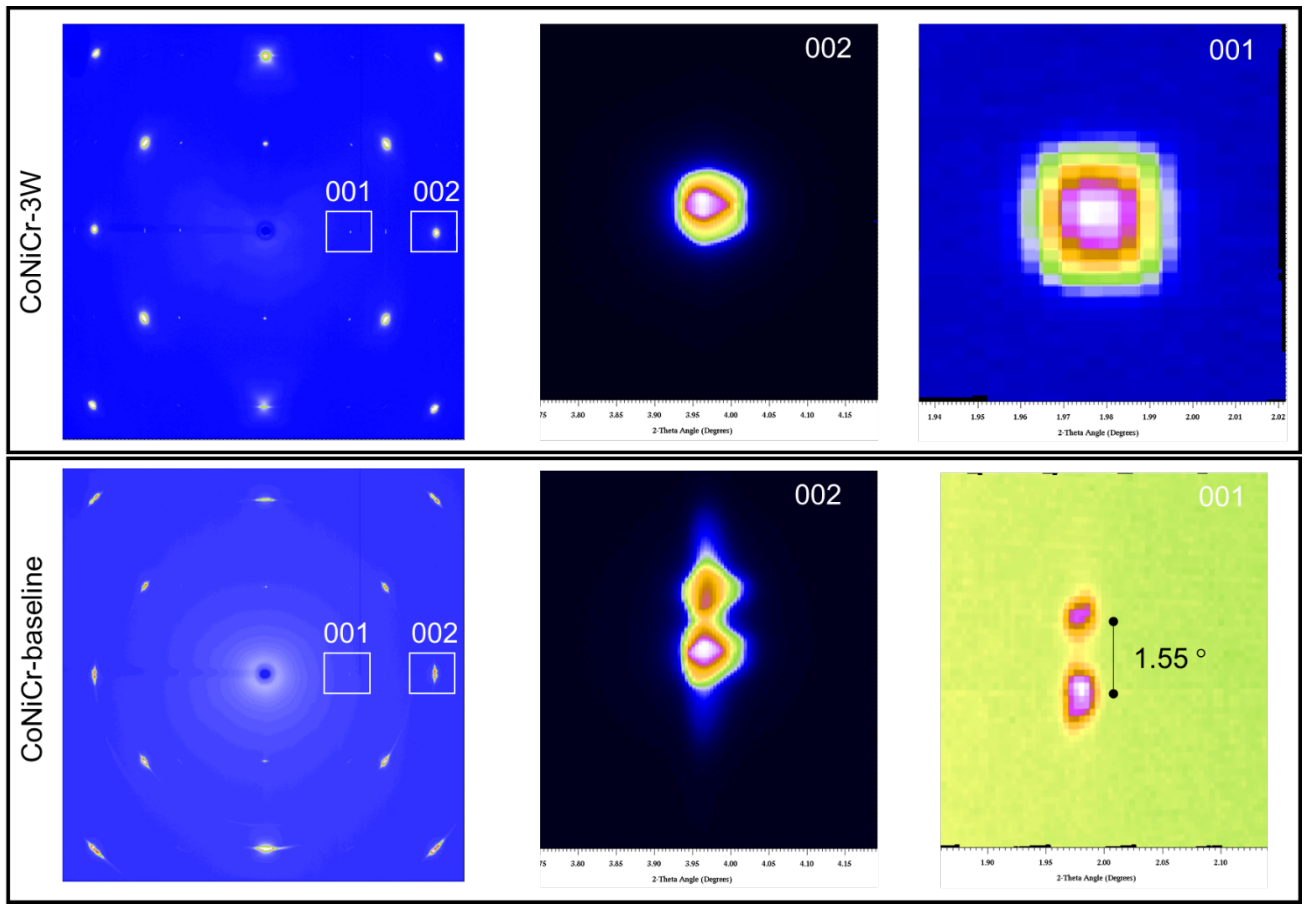

**Fig. S2** Comparison of diffraction spots obtained by synchrotron radiation HEXRD with a 360° scanning between a high-quality single crystal (CoNiCr-3W) and a defective single crystal (CoNiCr-baseline) containing subgrain. In the CoNiCr-3W single crystal, the (001) diffraction spot reflects its cubicity, aligning with the shape of γ′ precipitates. Conversely, the CoNiCr-baseline single crystal exhibits a distinct dumbbell character, indicating the presence of a subgrain within the sample (with a misorientation of 1.55° between the matrix grain and subgrain).

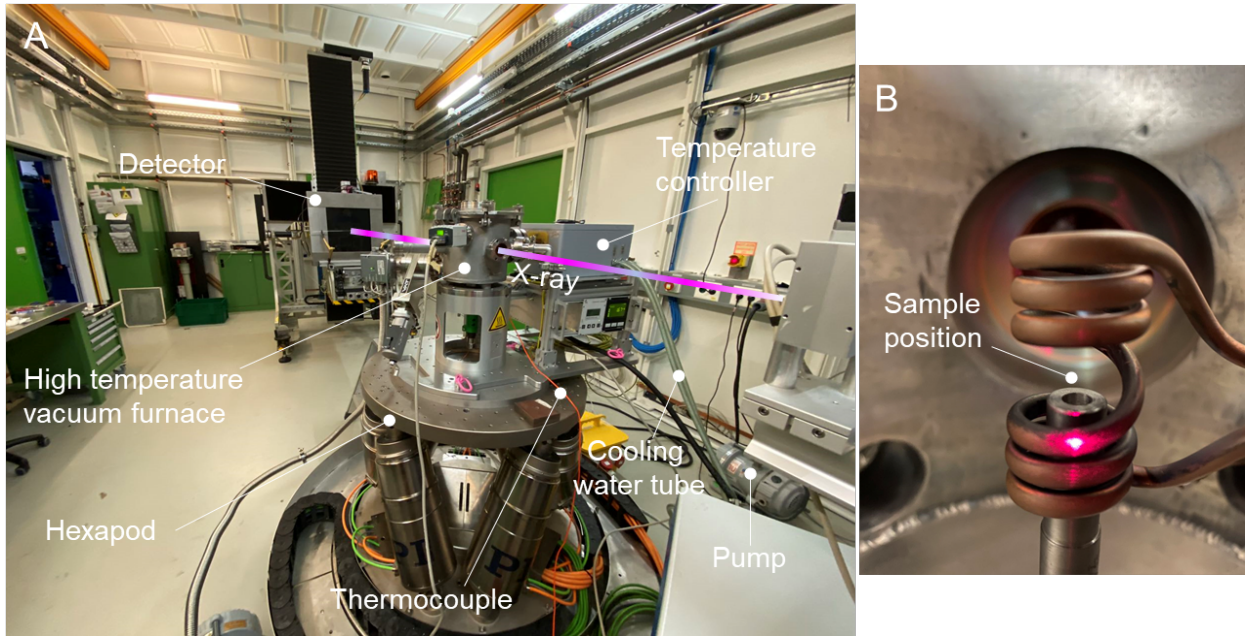

**Fig. S3** (A) Overview of the 3DXRD experimental setup based on hexapod. (B) The inside of the high temperature vacuum furnace. The sample is mounted on a sample stage made of Nb with high melting point. The coils are induction coil for heating sample. The high temperature vacuum furnace is fixed on the hexapod that allows x, y, and z translations as well as a rotation of the angle ω about the beam direction.

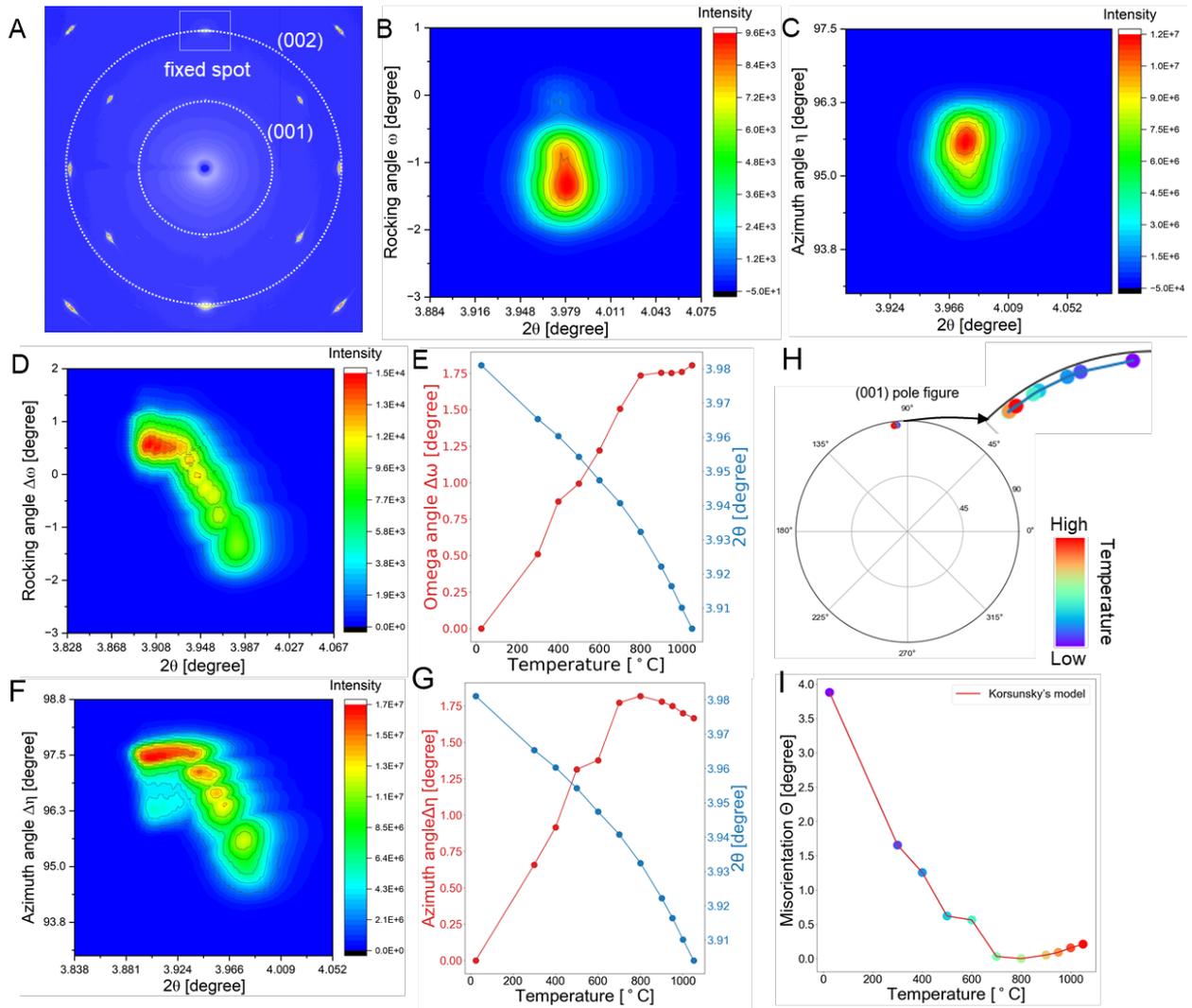

**Fig. S4 Direct observation of crystal rotation in single crystal CoNiCr-baseline.** (A) The diffraction spots obtained by 3DXRD with a ±180° scanning (φ), representing a single crystal (CoNiCr-baseline). (B) The contour map of reflection spots (002) with varying rocking angles (ω). (C) The contour map of reflection spots (002) with different azimuth angles (η). (D) The rocking angle (Δω) and diffraction angle (2θ) diffraction spots (002) at different temperatures. (E) Spatial positions for rocking angle (Δω) and diffraction angle (2θ) of the spots based on the center-of-mass (COM) method, singled out in **Fig. S4(D)**. (F) The summation of azimuth angle (Δη) and diffraction angle (2θ) reflection spots at different temperatures. (G) Spatial positions for azimuth angle (Δη) and diffraction angle (2θ) of the reflection based on the COM method, singled out in **Fig. S4(F)**. (H) The stereographic projection pole figure of crystal at different temperatures, based on Korsunsky's model. (I) The misorientation of the single crystal at different temperatures (the temperature 800 °C as reference case).

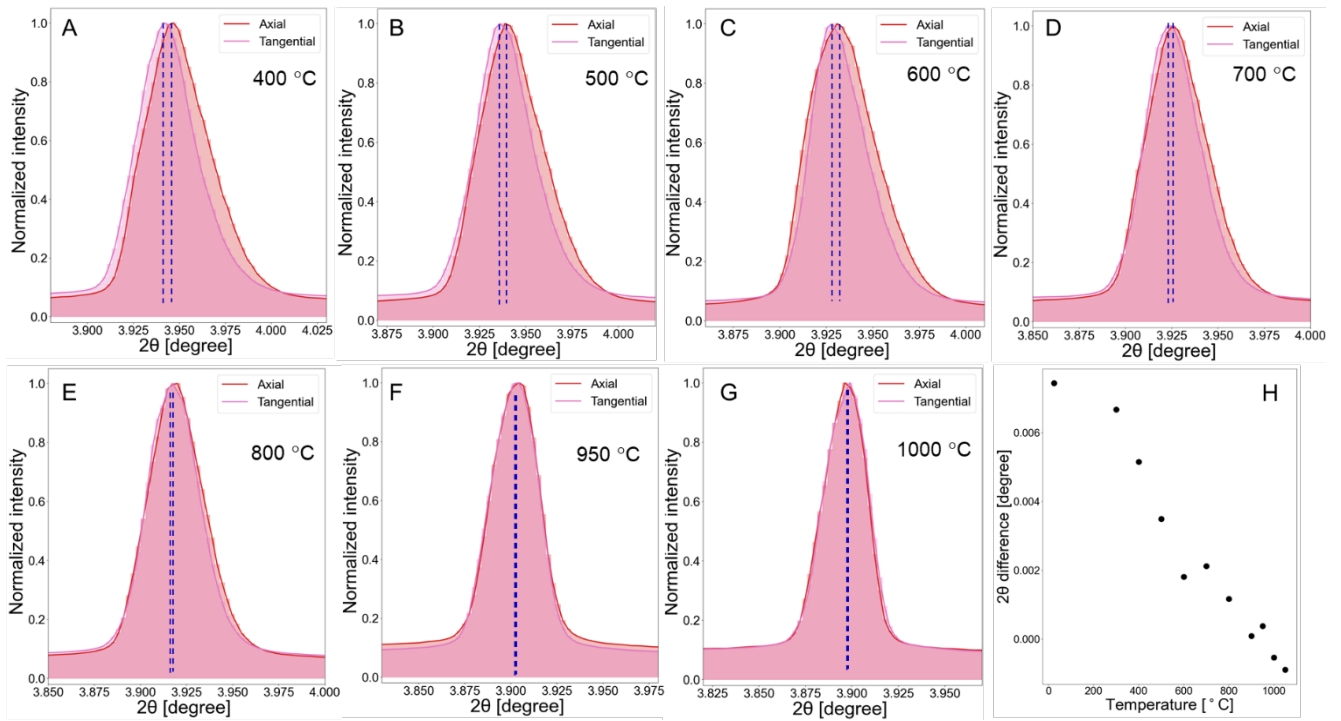

**Fig. S5** The line profiles with the intensity-diffraction angle (2θ) as a function of the temperature in axial and radial directions, respectively and the diffraction angle (2θ) difference at different temperatures in axial and tangential directions. This figure is complementary to **Fig. 3**.

**Table S1** The angles (θ, ω, η,) determined by the center-of-mass method, the ($\alpha_E$, $\beta_E$) polar angles and ($\phi$, $\varphi_1$, $\varphi_2$) Euler angles calculated based on Korsunsky's model[10] at different temperatures. All angles are given in degrees.

| Temperature / °C | θ | ω | η | $\varphi_1$ | $\phi$ | $\varphi_2$ | $\alpha_E$ | $\beta_E$ |
|---|---|---|---|---|---|---|---|---|
| 25 | 1.985 | 1.985 | -88.143 | -88.143 | 88.428 | 0.000 | 88.428 | -88.143 |
| 300 | 1.977 | 1.864 | -87.484 | -87.504 | 88.414 | 0.000 | 88.414 | -87.504 |
| 400 | 1.974 | 1.811 | -87.132 | -87.160 | 88.404 | 0.000 | 88.404 | -87.160 |
| 500 | 1.971 | 1.717 | -86.973 | -87.005 | 88.401 | 0.000 | 88.401 | -87.005 |
| 600 | 1.968 | 1.700 | -86.871 | -86.903 | 88.400 | 0.000 | 88.400 | -86.903 |
| 700 | 1.965 | 1.656 | -86.769 | -86.802 | 88.400 | 0.000 | 88.400 | -86.802 |
| 800 | 1.960 | 1.598 | -86.704 | -86.740 | 88.401 | 0.000 | 88.401 | -86.740 |
| 900 | 1.956 | 1.562 | -86.667 | -86.703 | 88.402 | 0.000 | 88.402 | -86.703 |
| 950 | 1.953 | 1.570 | -86.675 | -86.712 | 88.406 | 0.000 | 88.406 | -86.712 |
| 1000 | 1.950 | 1.580 | -86.694 | -86.730 | 88.409 | 0.000 | 88.409 | -86.730 |
| 1050 | 1.946 | 1.588 | -86.710 | -86.746 | 88.413 | 0.000 | 88.413 | -86.746 |

**Table S2** The simulation parameters required, such as Young's modulus[25], Poisson ratio[26], heat conductivity[27], density[20] (8570 kg/m$^3$), thermal expansion coefficient[27] and yield stress[20] for simulations with the finite element method (FEM).

| Temperature / °C | Young's modulus / GPa | Poisson ratio | thermal expansion coefficient / K$^{-1}$ | yield stress / MPa | Temperature / °C | Heat conductivity / W/(m·K) |
|---|---|---|---|---|---|---|
| 25 | 222 | 0.3885 | 8.17E-06 | 712 | 18 | 9.094 |
| 100 | | | 9.51E-06 | | 98 | 10.679 |
| 200 | | | 1.13E-05 | | 199 | 12.2 |
| 300 | 209 | 0.3945 | 1.23E-05 | | 299 | 13.567 |
| 400 | 203 | 0.3945 | 1.32E-05 | | 401 | 15.25 |
| 500 | 196 | 0.395 | 1.34E-05 | | 500 | 16.61 |
| 600 | 187 | 0.3955 | 1.36E-05 | 556 | 599 | 19.855 |
| 700 | 178 | 0.397 | 1.39E-05 | | 701 | 21.714 |
| 750 | | | | 556 | 799 | 22.285 |
| 800 | 168 | 0.402 | 1.42E-05 | | 898 | 23.049 |
| | | | | 519 | 998 | 24.851 |
| 900 | 156 | 0.409 | | | 1097 | 26.418 |
| 1000 | 144 | 0.42 | 1.52E-05 | | 1197 | 27.645 |
| 1050 | 137 | 0.42 | 1.62E-05 | | | |

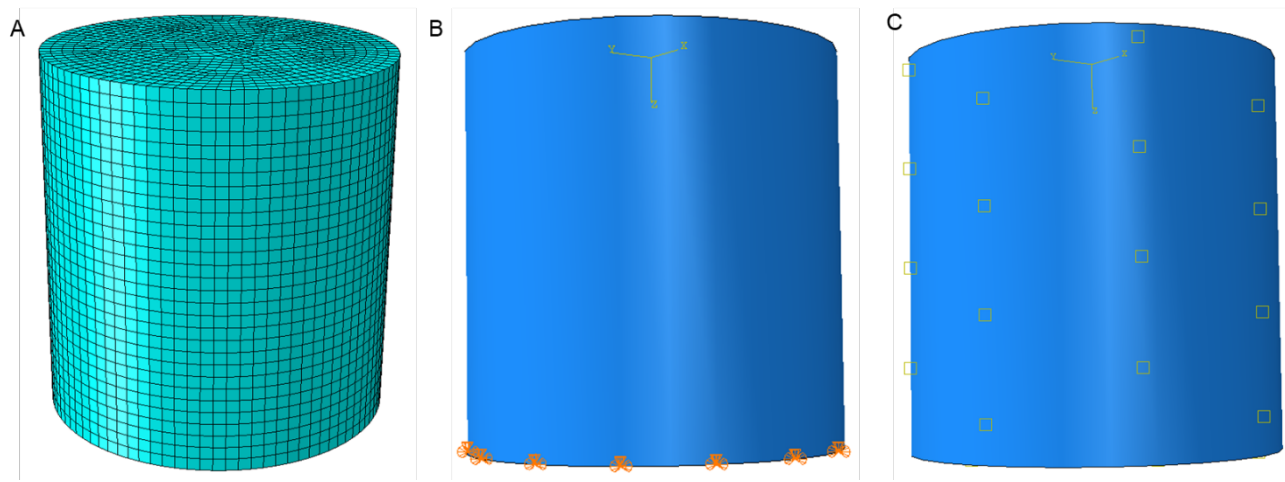

**Fig. S6** (A) The meshed cylinder for finite element method simulation. (B) The bottom plane was fixed as boundary conditions. (C) The interaction surface of the whole cylinder and air.